\documentclass{ws-p8-50x6-00}

\begin{document}

\title{\Large\bf 
Cosmology and ~~~~~~~~~~~~~~~~~~~~Modulus Stabilization  
~~~~~~~~~~~~~~~~~~~~in the Randall-Sundrum Setup 
~~~~~~~~~~~~~~~with Bulk Matter \footnote{
Proceeding of the talk at COSMO-2000, 
4-th International Workshop on Particle physics and the Early Universe, 
at Cheju-Island, Korea, Sept. 4-8, 2000.}} 

\author{
Bumseok Kyae}
\address{ Department of Physics and Center for Theoretical
Physics, Seoul National University,
Seoul 151-742, Korea \\
E-mail: kyae@fire.snu.ac.kr}
\maketitle

\abstracts{ 
We provide the exact time-dependent cosmological solutions in the 
Randall-Sundrum (RS) setup with bulk matter, which may be smoothly 
connected to the static RS metric.
In the static limit of the extra dimension, 
the solutions are reduced to the standard Friedmann equations. 
In view of our solutions, 
we also propose an explanation for how the extra 
dimension is stabilized in spite of a flat modulus potential
at the classical level.  }

As a possible solution of the gauge hierarchy problem, 
Randall and Sundrum (RS) proposed an $S^1/Z_2$ orbifold model 
with non-factorizable geometry of space-time 
\cite{rs}, which has immediately attracted a great deal of attention.    
The model employs two branes, Brane 1 (B1) with a positive cosmological
constant(or brane tension) $\Lambda_1\equiv 6k_1M^3$ and Brane 2 (B2) 
with a negative cosmological constant $\Lambda_2\equiv 6k_2M^3$, 
and introduces a negative bulk cosmological 
constant $\Lambda_b\equiv -6k^2M^3$. 
B1 is interpreted as the hidden brane and B2 is
identified with the visible brane \footnote{
Although B1 is regarded as the visible brane, 
the hierarchy problem between two scales could be solved 
by introducing bulk messenger fields and SUSY~\cite{mu}.  } 
Then the metric has an exponential warp factor which could be
used to understand the huge gap between the Planck and eletroweak scales.
Although the RS setup introduces cosmological constants $k$ in the bulk and
$k_1$ and $k_2$ on the
branes, it still describes a static universe because of
the fine-tuning between
the bulk and brane cosmological constants $k=k_1=-k_2$, which is a
consistency condition in the model.  Hence, if the fine-tuning is not
exact, the solution has the time dependence and the universe
expands exponentially~\cite{nihei} but its form is not suitable for 
the standard Big Bang universe after the inflation. 

Although there have been many cosmological solutions in the RS 
setup~\cite{cline,cosmo}, 
the graceful exit problem from the inflation phase to
the standard Big Bang cosmology has not been seriously considered
yet. In addition, the role of the extra dimension in the presence
of the bulk matter is not well understood.
In this talk
~\cite{original}, we will present exact cosmological solutions 
in the RS setup with bulk matter.  
Our exact solutions converge to the RS metric if the space time is made 
to be static, and leads to the standard Friedmann equations  
if the fifth dimension is stabilized.
In view of our exact solutions, we can find a clue
for a stabilization mechanism of the fifth dimension
and obtain a small compactified fifth dimension naturally.

Throughout this talk we consider a (4+1) dimensional universe with 
coordinate indexed by (0, 1, 2, 3, 5).  
The action describing the bulk matter as well as bulk gravity 
and brane matter is
\begin{equation}
\label{action}
S=\int d^5x\sqrt{-g}\left( {R\over 2}-\Lambda_b +{\cal L}^{(M)} \right)
+\sum_{j=1,2\ {\rm branes}}\int d^4x\sqrt{-g^{(j)}}\left({\cal L}^{(M)}_j
- \Lambda_j\right)~,
\end{equation}  
where we set the fundamental scale $M=1$.
${\cal L}^{(M)}$ and ${\cal L}^{(M)}_j$ represent 
matter contributions in the bulk and 
on the branes. For compatibility with the cosmological principle 
that our three dimensional space is homogeneous and isotropic, 
we assume that the metric of the universe has the following form,
\begin{equation}\label{metric}
ds^2
=-e^{2N(\tau,y)}d\tau^2+e^{2A(\tau,y)}
\delta_{ij}dx^idx^j+e^{2B(\tau,y)}dy^2~,
\end{equation}
where $\tau$ denotes time and $y$ denotes the fifth component. From 
the metric ansatz, the Einstein tesor $G_{MN}$ is derived through 
the standard calculation,
\begin{eqnarray}
\label{Ee:00} 
&&G^{(1)}\,_{00}=-3e^{2(N-B)}\left[A''+2A'^2-A'B'\right]
\\
&&G^{(2)}\,_{00}=3\left[\dot{A}^2+\dot{A}\dot{B}\right]
\\
\label{Ee:ii}
&&G^{(1)}\,_{ii}=e^{2(A-B)}\left[2A''+3A'^2+N''+N'^2+2A'N'-2A'B'-B'N'\right]
\\
&&G^{(2)}\,_{ii}=-e^{2(A-N)}\left[2\ddot{A}+3\dot{A}^2+\ddot{B}+\dot{B}^2
+2\dot{A}\dot{B}-2\dot{A}\dot{N}-\dot{B}\dot{N}\right]
\\
\label{Ee:55}
&&G^{(1)}\,_{55}=3\left[A'^2+A'N'\right]
\\
&&G^{(2)}\,_{55}=-3e^{2(B-N)}\left[\ddot{A}+2\dot{A}^2-\dot{A}\dot{N}\right]
\\
\label{Ee:05}
&&G_{05}=3\left[A'\dot{B}+\dot{A}N'-\dot{A}'-\dot{A}A'\right]~,
\end{eqnarray}
where dot and prime denote the derivatives with respect to 
$\tau$ and $y$, respectively, and the $i$ runs through $1$, $2$, 
and $3$. Here diagonal Einstein tensors are split into two parts,
$G^{(1)}\,_{AA}$ and $G^{(2)}\,_{AA}$, depending on the nontrivial $y$
and $\tau$ derivatives, respectively.   
Thus the original Einstein tensor is, of course, expressed as the sum, 
$G_{AA}\equiv G^{(1)}\,_{AA}+G^{(2)}\,_{AA}$, where the $A$ is $(0,i,5)$. 
The source part of the Einstein equation is composed of 
the cosmological constant and the energy-momentum tensor of matter.  
In this talk, we will regard the matter as perfect fluid.  
For the future convenience, we divide also the source tensor into two parts,
\begin{eqnarray}
T^{(1)}\,^A\,_B&=&-(1-\eta)
\cdot{\rm diag}\left[\Lambda_b, \Lambda_b, \Lambda_b, 
\Lambda_b, \Lambda_b\right] \nonumber \\
&&-\sum_{j=1,2\ {\rm branes}}\delta(y-y_j)e^{-B}{\rm diag}
\left[\Lambda_j, \Lambda_j, \Lambda_j, \Lambda_j, 0\right] \\ 
&&+\sum_{j=1,2\ {\rm branes}}\delta(y-y_j)e^{-B}{\rm diag}\left[-\hat{\rho}_j,
\hat{p}_j,\hat{p}_j,\hat{p}_j,0\right]\nonumber
\\
T^{(2)}\,^A\,_B &=& {\rm diag}\left[-(\rho+\eta\Lambda_b), P-\eta\Lambda_b, 
P-\eta\Lambda_b, P-\eta\Lambda_b, P_5-\eta\Lambda_b\right]~,
\end{eqnarray}
where the $\hat{\rho}_j$ and $\hat{p}_j$ are nontrivial components of 
the energy-momentum tensor of the matter living only on the $j$-th brane,
and $\eta$ is a number representing how $\Lambda_b$ is split
into $T^{(1)}\,^A\,_B$ and $T^{(2)}\,^A\,_B$. 
The total source tensor is described as 
$T^A\,_B=T^{(1)}\,^A\,_B+T^{(2)}\,^A\,_B$. 
Here we set $T_{05}=0$ because it is believed that there is no 
flow of matter along the fifth direction.  
The continuity equation of the energy-momentum tensor
$T^A\,_{B\ ;A}=0$ must be satisfied, whose
$B=0$ and $B=5$ components are 
\begin{eqnarray}
&&\dot\rho+3\dot{A}\left( \rho +P\right)+\dot{B}\left( \rho+ P_5 \right) = 0 
 \label{fluid1}\\
&&P'_5 + 3A'\left( P_5 - P\right) + N'\left(\rho+P_5 \right) = 0~.
 \label{fluid2}
\end{eqnarray}
The $B=i$ component is identically zero.  

Now let us take some ansatze,
\begin{eqnarray}
G^{(1)}\,_{AA}&=&T^{(1)}\,_{AA} \ \ \ ({\rm or\ \ } 
G^{(2)}\,_{AA}=T^{(2)}\,_{AA})\label{ans1} \\
A'(\tau,y)&=&N'(\tau,y)~.\label{ansatz}
\end{eqnarray}  
The above ansatze have been chosen to fulfill our purpose of restoring 
the Randall-Sundrum metric in the static limit.  
The ansatz Eq.~(\ref{ans1}) and $G_{05}$ read  
\begin{eqnarray}
3e^{-2B}&&\left[A''+2A'^2-A'B'\right] \label{00}\\
&&=-(1-\eta)\Lambda_b-e^{-B}\left[\delta(y)\left(\Lambda_1+\rho_1\right)
+\delta(y-1/2)\left(\Lambda_2+\rho_2\right)\right]
\nonumber \\
e^{-2B}&&\left[2A''+3A'^2+N''+N'^2+2A'N'-2A'B'-B'N'\right] \label{ii}\\
&&=-(1-\eta)\Lambda_b-e^{-B}\left[\delta(y)\left(\Lambda_1-p_1\right)
+\delta(y-1/2)\left(\Lambda_2-p_2\right)\right]
\nonumber \\
3e^{-2B}&&\left[A'^2+A'N'\right]=-(1-\eta)\Lambda_b
\label{55}\\
A'\dot{B}+&&\dot{A}N'-\dot{A}'-\dot{A}A'=0 ~~. \label{05}
\end{eqnarray}
Under the ansatz Eq.~(\ref{ansatz}), Eq.~(\ref{55}) becomes 
\begin{equation}
A'^2=-(1-\eta)\frac{\Lambda_b}{6}\times e^{2B} \equiv k^2e^{2B}~. 
\end{equation}
The solution consistent with the $S^1/Z_2$ orbifold symmetry is   
\begin{equation} \label{rssol}
A(\tau,|y|)\equiv k F(\tau,|y|)+J(\tau) ~~~{\rm and}~~~ 
F(\tau,|y|)'=-e^{B(\tau,|y|)}sgn(y) ~, 
\end{equation}
where the $sgn(y)$ is defined as 
$sgn(y)\equiv |y|'=2[\theta(y)-\theta(y-1/2)]-1$.  
Then, because of the ansatz Eq.~(\ref{ansatz}), 
the exponential factor $N$ of the $g_{00}$ component in our metric tensor 
is written as 
\begin{equation}
N(\tau,|y|)=k F(\tau,|y|)+K(\tau) \longrightarrow k F(\tau,|y|), 
\label{nrssol}
\end{equation}
where $K(\tau)$ is removed by the redefinition of 
time $\tau$ in the second part of the above equation.  
Therefore, we ignore $K(\tau)$ below.  

The above result Eq.~(\ref{rssol}) leads to
some useful relations,
\begin{eqnarray}
A''&=&-ke^{B}B'sgn(y)-2\left[\delta(y)
-\delta(y-1/2)\right]ke^{B}\nonumber 
\\
&=&A'B'-2\left[\delta(y)-\delta(y-1/2)\right]ke^{B} \label{rel1}\\
\dot{A}'&=&-ke^{B}sgn(y)\dot{B}=A'\dot{B} ~~. \label{rel2}
\end{eqnarray}
Eq.~(\ref{rel2}) implies that the $N(\tau,|y|)$ should be stabilized 
if the $B(\tau,|y|)$ can be stabilized somehow, since $A'=N'$.  
With Eqs.~(\ref{ansatz}) and (\ref{rel2}), we can show directly that 
our ansatz is consistent with Eq.~(\ref{05}).   
Because of Eqs.~(\ref{ansatz}) and (\ref{rel1}), Eqs.~(\ref{00}) 
and (\ref{ii}) just require matching the boundary conditions, 
\begin{equation} \label{branematter}
k=\frac{1}{6}\left(\Lambda_1+\hat{\rho}_1\right)
=-\frac{1}{6}\left(\Lambda_2+\hat{\rho}_2\right)~~~{\rm and}~~~
\hat{\rho}_j=-\hat{p}_j~.
\end{equation}  
Hence, considering the fluid continuity equation on the branes,  
$\dot{\hat{\rho}}_j+3\dot{A}(\hat{\rho}_j+\hat{p}_j)=0$, 
we can arrive at a result $\hat{\rho}_j=constant$ and so the $\hat{p}_j$ 
is a constant also, which are expected results 
from our assumption $T_{05}=0$.   

Now that we have fulfilled the ansatz Eq.~(\ref{ans1}) already, the  
remaining equations, $G^{(2)}\,_{AA}=T^{(2)}\,_{AA}$ are  
\begin{eqnarray}
\rho+\eta\Lambda_b&=&3e^{-2N}\left[\dot{A}^2+\dot{A}\dot{B}\right]
\label{rho}\\
P-\eta\Lambda_b&=&-e^{-2N}\left[2\ddot{A}+3\dot{A}^2+\ddot{B}+\dot{B}^2
+2\dot{A}\dot{B}-2\dot{A}\dot{N}-\dot{B}\dot{N}\right]
\label{p}\\
P_5-\eta\Lambda_b
&=&-3e^{-2N}\left[\ddot{A}+2\dot{A}^2-\dot{A}\dot{N}\right]~,  \label{p5} 
\end{eqnarray}
which describe the relation between matter and geometry dynamics.  
They are nothing but the extended Friedmann equations.  
With Eqs.~(\ref{ansatz}) and (\ref{rel2}), we can check that 
the above equations Eqs.~(\ref{rho}), (\ref{p}) and (\ref{p5}) satisfy 
both fluid continuity equations, 
Eqs.~(\ref{fluid1}) and (\ref{fluid2}) identically,  
that is, {\it the constraints, Eqs.~(\ref{fluid1}) and (\ref{fluid2}) are 
just redundant equations}, which are interesting results.
Therefore, the remaining required conditions for the solution are 
only Eqs.~(\ref{rssol}) and (\ref{nrssol}).  
The relations among the $\rho$, $P$ and $P_5$ may be, of course, governed by
particle physics.  
 
Toward a simple solution, let us consider the case that 
the size of the extra dimension is stabilized, i.e. $\dot{B}=0$, which leads 
also to $B'=0$ generically by redefinition of  $y$.  
Because of Eqs.~(\ref{ansatz}), (\ref{nrssol}) and (\ref{rel2}), then, 
$\dot{N}$ is also generically set to zero.  After all we have      
\begin{equation}
\dot{B}=B'=\dot{N}=0~. \label{staticcondi}
\end{equation}
Then, from Eqs.~(\ref{rssol}) and (\ref{nrssol}), 
the function $F(\tau,|y|)$ is determined to $F(\tau,|y|)=-ke^{B}|y|$ and so
\begin{eqnarray}
N(\tau,|y|)&=&-ke^B|y|\equiv -kb_0|y|~~~ \\
A(\tau,|y|)&=&-kb_0|y|+J(\tau)\equiv -kb_0|y|+\int ^{\tau}H(t)dt ~,
\end{eqnarray}
where the interval scale $b_0$ is a small constant and $H(\tau)$ is a 
time dependent arbitrary function but may be determined by the equation of 
state.  
Thus the metric is read off as 
\begin{equation}
ds^2=e^{-2kb_0|y|}\left(-d\tau^2+e^{2\int ^{\tau}H(t)dt}d\vec{x}^2
\right)+b_0^2dy^2~~.
\end{equation}
Note that the metric is the same as that of the Randall-Sundrum except for the 
factor $e^{2\int^{\tau}H(t)dt}$. Then   
Eqs.~(\ref{rho}), (\ref{p}) and (\ref{p5}) become
\begin{eqnarray}
\rho(\tau,|y|)+\eta\Lambda_b&=&3e^{2kb_0|y|}H^2(\tau) \label{friedmann} \\
P(\tau,|y|)-\eta\Lambda_b
&=&-e^{2kb_0|y|}\left[2\dot{H}(\tau)+3H^2(\tau)\right] \label{PH}\\
P_5(\tau,|y|)-\eta\Lambda_b
&=&-3e^{2kb_0|y|}\left[\dot{H}(\tau)+2H^2(\tau)\right]\nonumber \\
&=&-\frac{1}{2}\bigg[\rho(\tau,|y|)-3P(\tau,|y|)\bigg]-2\eta\Lambda_b
\label{fp5} \\ &=&\frac{1}{2}T^{(2)\mu}\,_{\mu} ~, \nonumber
\end{eqnarray}
where $\mu$ runs through 0, 1, 2, 3.  
[$B(\tau,|y|)$ is associated with the vacuum expectation 
value of a massless four-dimensional scalar field.]   
The above equations, Eqs.~(\ref{friedmann}), (\ref{PH}) 
and (\ref{fp5}), show that due to the exponential factor 
{\it matter in the bulk is 
accumulated mainly near the B2 brane (negative tension brane)}.  

Of course, any $H(\tau)$ with 
{\it $H(\tau)\rightarrow 0$ and $\dot{H}(\tau)\rightarrow 0$ 
as $\tau \rightarrow \infty$ can lead to an exit from 
an inflationary phase to a static Randall-Sundrum ($k\neq 0$) 
or Minkowski ($k=0$) universe}.  
In this talk, however, we will not specify a model because we are more 
interested in the real {\it expanding} universe. 

In Eqs.(\ref{friedmann}), (\ref{PH}) and (\ref{fp5}), we should remember that 
$\rho+\eta\Lambda_b$, $P-\eta\Lambda_b$ and $P_5-\eta\Lambda_b$ are 
non-trivial components of the five-dimensional energy-momentum tensor.  
To derive effective four-dimensional energy-momentum tensor $\tilde{T}^A\,_B$, 
it is necessary to consider the definition of the five-dimensional 
energy-momentum tensor, 
\begin{equation}
\delta S = \int d^5x\sqrt{-g}~\delta(g^B\,_A)T^{(2)A}\,_B
=\int d^4x \int dy ~b_0\sqrt{-g_4}~\delta(\delta^B\,_A)T^{(2)A}\,_B~,
\end{equation}
where $g_4\equiv {\rm det}[g_{\mu \nu}]$ ($\mu ,\nu=0,1,2,3$).  
As four-dimensional metric at a four-dimensional slice in the bulk is 
$\tilde{g}_{\mu \nu}=e^{2kb_0|y|}g_{\mu \nu}$, which was introduced 
by Randall and Sundrum to solve the gauge hierarchy problem \cite{rs}, 
effective four-dimensional energy-momentum tensor is given as
\begin{eqnarray}
&&\tilde{T}^{\mu}\,_{\nu}=b_0\int dy~e^{-4kb_0|y|}T^{(2)\mu}\,_{\nu} 
\nonumber \\ \label{4dfried}
&&=-b_0\int dy~e^{-2kb_0|y|} \nonumber \\
&&~\cdot {\rm diag}[3H^2(\tau),
2\dot{H}(\tau)+3H^2(\tau),2\dot{H}(\tau)+3H^2(\tau),
2\dot{H}(\tau)+3H^2(\tau)] \\ \label{4dfried2}
&&\equiv{\rm diag}[-(\tilde{\rho}(\tau)+\eta \tilde{\Lambda}),
\tilde{P}(\tau)-\eta \tilde{\Lambda},\tilde{P}(\tau)-\eta \tilde{\Lambda},
\tilde{P}(\tau)-\eta \tilde{\Lambda}]~~.
\end{eqnarray}  
According to RS, $b_0\int dy e^{-2kb_0|y|}$ is nothing but 
the induced four-dimensional Planck scale $M_{Pl}^2$ \cite{rs}.  
Thus, from Eqs.(\ref{friedmann}), (\ref{PH}), (\ref{4dfried}) 
and (\ref{4dfried2}), we can get the Friedmann equations, 
\begin{eqnarray} \label{standardfried} 
\bigg[\frac{\dot{a}(\tau)}{a(\tau)}\bigg]^2&=&
\frac{e^{-2kb_0|y|}}{3}\bigg[\rho(\tau,|y|)+\eta\Lambda_b\bigg]
=\frac{1}{3M_{Pl}^2}\bigg[\tilde{\rho}(\tau)+\eta\tilde{\Lambda}\bigg] 
~~~{\rm and}~~~\\
\frac{\ddot{a}(\tau)}{a(\tau)}~~&=&
-\frac{e^{-2kb_0|y|}}{6}\bigg[\rho(\tau,|y|)+3P(\tau,|y|)
-2\eta\Lambda_b\bigg] \nonumber \\
&=&-\frac{1}{6M_{Pl}^2}\bigg[\tilde{\rho}(\tau)+3\tilde{P}(\tau)
-2\eta\tilde{\Lambda}\bigg]~,
\end{eqnarray}
where $a(\tau)$ is a scale factor of our three-dimensional space,
$a(\tau)\equiv e^{\int^{\tau} H(t)dt}$.
Therefore, {\it we could have saved 
the whole standard cosmological scenario
by only requiring stabilization of 
$B$ in our framework (i.e. $\dot B=0$)}.  

As an example, let us consider the vacuum dominated era.  
The equation of state is $P=-\rho$  
and then $P_5$ is given by $P_5=-2\rho-\eta \Lambda_b$.  
Then Eqs.~(\ref{friedmann}) and (\ref{PH}) lead to 
\begin{equation}
H=constant\equiv H_0~, 
\end{equation}
which gives the inflationary universe. 
$H_0$ can be considered as a parameter representing the degree of  
fine-tuning given in Eq.~(\ref{friedmann}). If $H_0$ vanishes,
the fine-tuning is successful and the universe 
is static.  On the other hand, a non-zero $H_0$ does not satisfy
the fine-tuning condition and gives rise to an inflationary universe.
Note that our solution is for $\dot{B}=0$ and 
any modulus potential is not generated at the classical level 
since $b_0$ is arbitrary.  

We proceed to discuss the possibility of stabilizing the size of the fifth 
dimension in our framework. Since we obtained already 
the solutions for the $\dot{B}=0$ case, we conclude 
that if $\dot{B}$ goes to zero asymptotically, there exist 
solutions converging asymptotically to our $\dot{B}=0$ solutions.

We suppose that the stabilization era occurs before or during
the conventional inflation era. Thus, we suppose that the system 
is governed by the same equations of state as those of the inflation 
era discussed above.  
Then, from Eqs.~(\ref{rho}), (\ref{p}) and (\ref{p5}), we obtain 
\begin{eqnarray}
2\ddot{A}+\ddot{B}+\dot{B}^2-\dot{A}\dot{B}-&&2\dot{A}\dot{N}-\dot{B}\dot{N}=0 
\label{stab1} \\
\ddot{A}-\dot{A}\dot{N}&&-2\dot{A}\dot{B}=0 ~, \label{stab2}
\end{eqnarray} 
which are the equations of state in the inflationary era.  
The above equations and Eqs.~(\ref{rssol}) and (\ref{nrssol}) convince us that 
any potential for the modulus field is not generated also 
since the $B$ is not fixed yet.  

Eq.~(\ref{stab2}) is easily solved, 
\begin{equation}
\dot{A}=e^{s(|y|)+N}b^2=e^{s(|y|)+kF}b^2~  
\bigg(=k\dot{F}+\dot{J}\bigg), \label{inca}  
\end{equation}
where $s(|y|)$ is an integration constant and $b$ is defined as 
$b(\tau,|y|)\equiv e^{B(\tau,|y|)}$.  
Note that $A$ is an increasing function of time since $\dot{A}>0$.  
Removing $\ddot A$ from Eqs.~(\ref{stab1}) and (\ref{stab2}),
we obtain 
\begin{equation}
\ddot{B}+\dot{B}^2+3\dot{A}\dot{B}-\dot{B}\dot{N}=0,
\end{equation}
which can be solved to give
\begin{equation}
\dot{b}=\frac{1}{e^{t(|y|)-N+3A}}=\frac{1}{e^{t(|y|)+2kF+3J}}~,  
\label{incb}
\end{equation}
where $t(|y|)$ is an integration constant.  
Note that $b$ is an increasing function of time but $\dot{b}$ 
could be zero asymptotically. The extra dimension scale 
$b$ can be stabilized if the combination $-N+3A~(=2kF+3J)$ is an 
increasing function of time without limit. 
Especially, if $N\ll A$ and
$A$ increases as $\tau\rightarrow\infty$ 
without a limit at least with a power law, which {\it inflates the
three dimensional space, then $\dot{b}$ would 
decrease exponentially and so $b$ could 
be stabilized.} Below we show that it is possible.   

$^{}$ From Eqs.~(\ref{inca}) and (\ref{incb}), we obtain
\begin{equation}
\frac{\dot{A}}{\dot{b}}=e^{s(|y|)+t(|y|)+3A}\ b^2 ~,
\end{equation} 
which is integrable. The solution is
\begin{equation}
b(\tau,|y|)^3=u^3(|y|)-e^{-s(|y|)-t(|y|)-3A} ~, \label{exactb}
\end{equation}
where $u(|y|)$ is a $|y|$ dependent arbitrary function.  
The remaining constraints to satisfy are only Eqs.~(\ref{rssol}) 
and (\ref{nrssol}), i.e. $A'=N'=-kb(\tau,|y|)sgn(y)$, from which 
Eq.~(\ref{exactb}) can be written as 
\begin{equation}
3\times \frac{u^2u^{\prime}-b^2b^{\prime}}{u^3(|y|)-b^3}
+s^{\prime}(|y|)+t^{\prime}(|y|)=-3A'=3kb\cdot sgn(y)~. \label{deu}
\end{equation} 
We intend to make our solution become the 
inflationary solution asymptotically that was obtained before.
Since $N\rightarrow -kb_0|y|$, $\dot{A}\rightarrow H_0$, and $b
\rightarrow b_0$ as $\tau \rightarrow \infty$, 
let us take the integration constants in Eqs.~(\ref{inca}) 
and (\ref{exactb}) as
\begin{eqnarray}
s(|y|)&=&kb_0|y|+\ln~H_0-2\ln~b_0 \label{H} \\ 
u(|y|)&=&b_0 ~,
\end{eqnarray}
where $b_0$ and $H_0$ were defined above.
So far the solutions were exact.
Note that any value for $H_0$ is possible, which 
does not influence Eq.~(\ref{deu}).  
Although Eq.~(\ref{deu}) is difficult to solve exactly,
we can argue that
$|N|=k|F|\ll \{J(\tau)~~{\rm and}~~ H_0\tau \}$
is sufficient to draw a meaningful conclusion.  
  
$J(\tau)$ and $H_0$ can be chosen always such that 
$|N|=k|F|\ll \{J(\tau)~~{\rm and}~~ H_0\tau \}$.  
It is compatible with the phenomenological requirement of a large
$H_0$ so that our solutions are derived during the inflationary epoch.
(Actually during the inflationary era, 
$H_0^{-1}$ is (a very large value)$^{-1}$ 
$\approx 10^{-34}$~sec.) 
In this {\it early} inflationary era, $N$ and $b$ are regarded as being 
much smaller than $H_0\tau$, since the scale of the universe was small
right after the Big-Bang. 
Then $-N+3A$ increases with time and 
the extra dimension scale $b$ is stabilized rapidly to $b_0$.  
Once $b$ is stabilized, the conditions Eq.~(\ref{staticcondi}) become 
valid and $N$ and $\dot{A}$ are forced to $-kb_0|y|$ and $H_0$, 
respectively.  

To see our argument explicitly, let us take somewhat large $J(\tau)$  
assumption and solve Eqs.~(\ref{inca}), (\ref{incb}) and (\ref{deu}) 
approximately under the condition $k|F|\ll J$.  Our aim is
to show $\dot A\rightarrow H_0$, $\dot b\rightarrow 0$, and 
$b\rightarrow b_0$.
Here, we set $kF$ and $b$ as $O(1)$ initially, and then $J\gg 1$.   
Then Eq.~(\ref{exactb}) gives 
\begin{equation} \label{appr}
b_0^3-b^3(\tau,|y|)=\frac{H_0}{b_0^2}e^{-kb_0|y|-3kF}e^{-3J}\ll 1~~.
\end{equation} 
With Eqs.~(\ref{deu}) and (\ref{appr}) we are led to the following results, 
\begin{equation}
b^{\prime}\ll 1~~~{\rm or}~~~b\approx b(\tau), 
\end{equation}
and we obtain
\begin{equation}
F(\tau,|y|)\approx -b(\tau)|y|~~. 
\end{equation}
Here we must set $t(|y|)\approx 0$ in view of Eq.~(\ref{incb}).  
Then, Eqs.~(\ref{inca}) and (\ref{incb})  
become 
\begin{equation}
\dot{J}\approx H_0\frac{b^2}{b_0^2}~~~{\rm and} ~~~\dot{b}\approx e^{-3J}
\ll 1,
\end{equation}
which shows that during the period of the three space inflation it
is quite difficult for $b$ to be dynamical. It is interpreted as the
stabilization of the extra dimension in spite of the flat potential
for $b$.   
$^{}$From Eq.~(52) we can derive an expression for $b(\tau)$, 
\begin{equation}
-\frac{1}{6b_0^2}\ln\bigg[\frac{(b_0-b)^2}{b_0^2+b_0b+b^2}\bigg]
+\frac{1}{b_0^2\sqrt{3}}\tan^{-1}\bigg[\frac{b_0+2b}{b_0\sqrt{3}}\bigg]
=\frac{H_0}{b_0^2}\tau, \label{k0}
\end{equation}
or
\begin{equation} 
b(\tau)\approx b_0-e^{-3H_0\tau}~.
\end{equation}
Here we can see that as $\tau \rightarrow \infty$, $b(\tau)$ grows to 
$b_0$ and $\dot{J}$ tends to $H_0$ asymptotically.  
In other words, {\it to obtain an inflationary universe 
$b$ should be stabilized to $b_0$ exponentially}.    
Note that Eq.~(\ref{k0}) becomes an {\it exact} result 
provided the warp factor vanishes, which corresponds the cases of 
$\Lambda_b=0$ or $\eta=1$.  

If $b$ is $O(1)$ but small right after the Big Bang, $b_0$ should be
$O(1)$ but small. As the extra dimension gets stabilized
($\dot B=0$) soon after the beginning 
of the inflationary era, while the three dimensional space inflates
(to $e^{H_0\tau}$), 
{\it $b$ remains small} ($\approx 1/M$) and 
the universe is reduced effectively to 4-dimension. 

In a similar method, we can show that $\dot{b}$ 
is made asymptotically to zero also in the radiation ($P=\rho/3$, $\eta=0$) 
and matter dominated era ($P=0$, $\eta=0$).  However, as the initial condition  
for $\dot{b}$ is zero (through the above solution in the
inflationary epoch), $b$ should have been stabilized already  
and so Eq.~(\ref{staticcondi}) should have been valid since the beginning of 
the radiation dominated era.  Therefore, the Friedmann equations 
Eqs.~(\ref{standardfried}) hold good in the radiation and matter dominated
eras.      
  
In conclusion, we have provided exact cosmological 
solutions in the RS setup with bulk matter.  
In the static limit of all components of 
the metric, the solutions become the 
RS metric and in static limit of the extra dimension, 
they are reduced to the standard Friedmann equations, 
which implies that bulk matter is accumulated mainly  
near the negative tension brane (visible brane B2).  
In this case the modulus potential is not generated effectively
at the classical level.  
With our solution, however, we have shown that the extra  
dimension could be stabilized (the $\dot B=0$ solution) 
even if the modulus potensial is flat ($b_0$ is arbitrary) and 
it should be small since the three dimensional space inflates
during the inflationary era.

\section*{Acknowledgments}
This work is based on the collaboration with J. E. Kim~\cite{original}, 
and is supported in part by the BK21 program of Ministry 
of Education.

\end{document}